\documentclass[twoside,11pt]{article}

\usepackage{jmlr2e}
\usepackage{multirow}
\usepackage{float}
\usepackage[table]{xcolor}
\usepackage{longtable,tabu}
\usepackage{siunitx}
\usepackage{multirow}
\usepackage{booktabs}

\definecolor{greend}{HTML}{118E1E}

\newcommand{\beginsupplement}{%
        \setcounter{table}{0}
        \renewcommand{\thetable}{S\arabic{table}}%
        \setcounter{figure}{0}
        \renewcommand{\thefigure}{S\arabic{figure}}%
     }

\firstpageno{1}

\begin{document}

\title{Self-supervised learning improves dMMR/MSI detection from histology slides across multiple cancers}

\author{\name Charlie Saillard \email charlie.saillard@owkin.com \\
      \addr Owkin, Inc.\\
      \AND
      \name Olivier Dehaene \email olivier.dehaene@gmail.com \\
      \addr Owkin, Inc.\\
      \AND
      \name Tanguy Marchand \email tanguy.marchand@owkin.com \\
      \addr Owkin, Inc.\\
      \AND
      \name Olivier Moindrot \email omoindrot@gmail.com \\
      \addr Owkin, Inc.\\
      \AND
      \name Aurélie Kamoun \email aurelie.kamoun@owkin.com \\
      \addr Owkin, Inc.\\
      \AND
      \name Benoit Schmauch \email benoit.schmauch@owkin.com \\
      \addr Owkin, Inc.\\
      \AND
      \name Simon Jegou \email simon.jegou@gmail.com \\
      \addr Owkin, Inc.}

\editor{}

\maketitle

\begin{abstract}
Microsatellite instability (MSI) is a tumor phenotype whose diagnosis largely impacts patient care in colorectal cancers (CRC), and is associated with response to immunotherapy in all solid tumors. Deep learning models detecting MSI tumors directly from H\&E stained slides have shown promise in improving diagnosis of MSI patients. Prior deep learning models for MSI detection have relied on neural networks pretrained on ImageNet dataset, which does not contain any medical image. In this study, we leverage recent advances in self-supervised learning by training neural networks on histology images from the TCGA dataset using MoCo V2. We show that these networks consistently outperform their counterparts pretrained using ImageNet and obtain state-of-the-art results for MSI detection with AUCs of 0.92 and 0.83 for CRC and gastric tumors, respectively. These models generalize well on an external CRC cohort (0.97 AUC on PAIP) and improve transfer from one organ to another. Finally we show that predictive image regions exhibit meaningful histological patterns, and that the use of MoCo features highlighted more relevant patterns according to an expert pathologist.
\end{abstract}


\section{Introduction}
Microsatellite Instability (MSI) is a frequent tumor phenotype characterized by an abnormal repetition of short DNA motifs caused by a deficiency of the DNA mismatch repair system (MMR). MMR deficient tumors (dMMR) result from defects in the major MMR genes, namely \textit{MLH1, MSH2, MSH6, PMS2}. These defects arise either sporadically or as a hereditary condition named Lynch syndrome (LS), predisposing patients to develop cancers in several organs.

Recent studies have shown that immune checkpoint blockade therapy has a promising response in dMMR/MSI cancers regardless of the tissue of origin [\cite{le2017mismatch}]. In 2017, this genomic instability phenotype became the first pan-cancer biomarker approved by the US FDA, allowing the use of pembrolizumab (Keytruda) for patients with dMMR/MSI solid tumors at the metastatic stage [\cite{prasad2018cancer}].

As of today, systematic MSI screening is only recommended for colorectal cancer (CRC) and endometrial cancer [\cite{svrcek2019msi}] where the prevalence is relatively high (10\% to 20\%), principally to detect LS patients and provide them with adequate follow-up. In early stages of CRC, MSI tumors are associated with good prognosis and resistance to chemotherapy [\cite{sargent2010defective}], making the diagnosis of this phenotype all the more essential for patient care and therapeutic decision. dMMR/MSI diagnosis is traditionally done using immunohistochemistry (IHC),  polymerase chain reaction (PCR) assays, or next generation sequencing. Those methods can be time-consuming, expensive, and rely on specific expertise which may not be available in every center.

Deep learning based MSI classifiers using H\&E stained digital images offer a new alternative for a broader and more efficient screening [\cite{echle2020clinical}]. In CRC, previous work suggests that the use of such models as pre-screening tools could eventually replace IHC and PCR for a subset of tumors classified as microsatellite stable (MSS) or unstable (MSI) with a high probability [\cite{kacew2021artificial}]. In other locations where MSI prevalence is lower and screening not done as routine practice, predictive models of MSI status from WSI could be used as efficient pre-screening tools.

In this work, we leverage recent advances in self-supervised learning (SSL) on images. We show that SSL permits to reach state-of-the-art results on colorectal and gastric cancer cohorts from The Cancer Genome Atlas (TCGA), generalizes well on an unseen colorectal cohort (PAIP), and could pave the way for classifiers on locations with low MSI prevalence.

\section{Related Work}

\paragraph{Expert models} Several histology patterns on H\&E images have been reported to correlate with MSI, such as tumor-infiltrating lymphocytes, lack of dirty necrosis or poor differentiation [(\cite{greenson2009pathologic}]. A series of models based on clinico-pathological features have been developed [\cite{greenson2009pathologic, jenkins2007pathology, hyde2010histology, fujiyoshi2017predictive, roman2010microsatellite}] and reported ROC-AUC performances ranging from 0.85 to 0.92 in various cohorts of patients with CRC. These methods however require time-consuming annotations from expert pathologists and are prone to inter-rater variability.

\paragraph{Deep learning models} In a seminal publication, \cite{kather2019deep} proved the feasibility to determine the dMMR/MSI status from H\&E stained whole slide images (WSI) using deep learning. They trained a first ResNet [\cite{he2016deep}] to segment tumor regions on WSI, and a second one to predict MSI/MSS status in each tumor tile. Each ResNet was pretrained on ImageNet and the last 10 layers were fine-tuned. Models were trained and validated on different TCGA cohorts and obtained respectively AUCs of 0.77
, 0.81
and 0.75
on Colorectal, Gastric and Endometrial formalin-fixed paraffin-embedded (FFPE) datasets. .

In a larger scale study focusing on CRC only [\cite{echle2020clinical}], the same team later trained a model on $n=6406$ patients, reaching 0.96 AUC (95\% CI  0.93–0.98) on an external dataset of $n=771$ patients. Tumor tissues were manually outlined by pathologists.

Since then, different works based on deep learning methods have been published [\cite{zhang2018adversarial, cao2020development, hong2020predicting, yamashita2021deep, bilal2021novel, lee2021feasibility}] and are reviewed in \cite{hildebrand2021artificial}. The vast majority rely on networks pretrained on the ImageNet dataset [\cite{deng2009imagenet}] and only the last layers are re-trained or fine-tuned. Most of these papers also rely on tumor segmentation as a first step in their models (either by a pathologist or by a deep learning model). 
\paragraph{Self-Supervised learning}
Over the past few years, rapid progress has been made in the field of SSL using contrastive learning or self-distillation strategies: simCLR [\cite{chen2020simple, chen2020big}] , MoCo [\cite{he2020momentum, chen2020improved, chen2021empirical}], BYOL [\cite{grill2020bootstrap}], achieving impressive performances on ImageNet without using any labels. Such models have also been shown to outperform supervised models in transfer learning tasks [\cite{chen2020simple, caron2021emerging, li2021efficient}].

These advances are of particular interest in medical imaging applications where labeled datasets are hard to collect, and especially in histology where each WSI contains thousands of unlabeled images. There is growing evidence that SSL is a powerful method to obtain relevant features for various prediction tasks from histology images [\cite{dehaene2020self, lu2019semi, li2021dual, koohbanani2021self, gildenblat2019self, abbet2021self, srinidhi2021self}]. In this work, we show  that self-supervision can be efficiently used to detect  dMMR/MSI tumors from histology slides, and outperform ImageNet pretrained models across and between several tumors.

\section{Methods}
\subsection{Proposed pipeline}
\label{subsection:pipeline}

\begin{figure}[!ht]
\centering
\includegraphics[width=16cm]{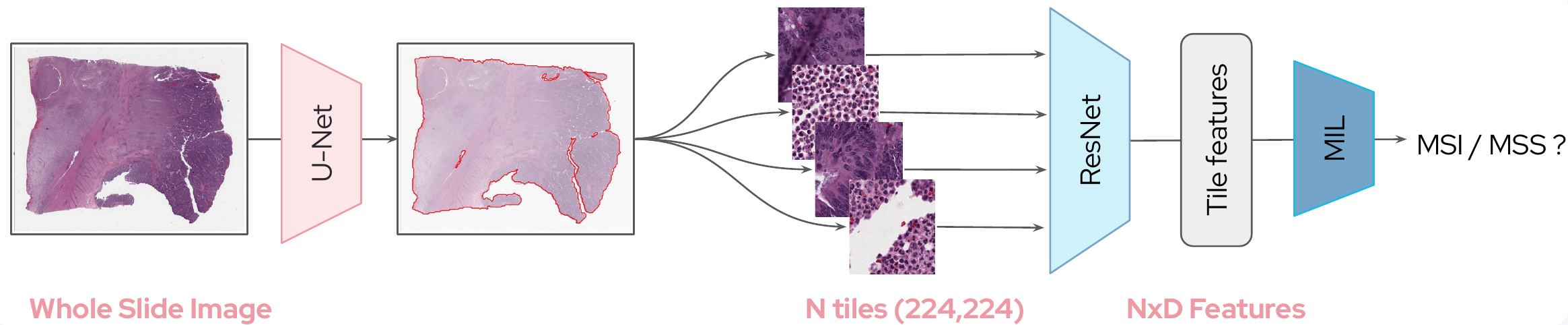}
\caption{Overview of the proposed pipeline. 
}
\label{fig:architecture}
\end{figure}

First, a U-Net neural network [\cite{ronneberger2015u}] is used to segment tissue on the input WSI and discard the background, as well as artifacts. Second, segmented tissue is divided into $N$ (typically between 10,000 and 30,000) smaller images called tiles. Each tile has a fixed shape of $224 \times 224$ pixels (resolution of 0.5 micron per pixel). Third, the $N$ tiles are embedded into feature vectors of shape $D$ using a pretrained convolutional neural network. Fourth, the $N\times D$ features are aggregated using a multiple instance learning model. This final model is the only one trained using MSI/MSS labels.

In this study, we benchmarked 2 different feature extractors (ResNet-50 pretrained with supervised learning on ImageNet [\cite{deng2009imagenet}] or with SSL on TCGA) and 3 multiple instance learning models (MeanPool, Chowder and DeepMIL).

\paragraph{ImageNet feature extraction} We first extracted features using the last layer of a ResNet-50 pretrained using supervised learning on the ImageNet-1k dataset. We used an auto-encoder to reduce dimension to $D=256$ because we observed empirically that it significantly improves the performances of Chowder while yielding similar results for MeanPool and DeepMIL. We did not observe similar improvements with MoCo features. 

\paragraph{MoCo feature extraction} Following \cite{dehaene2020self}, we trained several ResNet-50 models using Momentum Contrast v2 (MoCo v2[\cite{chen2020improved}]). We used the exact same parameters and data augmentation scheme but a bigger ResNet backbone was used (the bottleneck number of channels is twice larger in every block). Three different feature extractors were trained: MoCo-CRC using 4.7M tiles from TCGA-CRC, MoCo-Gastric using 4.7M tiles from TCGA-Gastric, and MoCo-CRC-Gastric using the concatenation of the two previous datasets. Each model was trained for 200 epochs (approximately 30 hours) on 16 NVIDIA Tesla V100. The obtained MoCo features have a dimension $D=2048$.

\paragraph{MeanPool} As a baseline multiple instance learning method, we used a simple average pooling of the tile features followed by a logistic regression either with or without $L2$ penalization ($C=0$, $0.5$ or $1$). 

\paragraph{Chowder} We implemented a variant of Chowder [\cite{courtiol2018classification}]. A multilayer perceptron (MLP) with 128 hidden neurons and sigmoid activation is applied to each tile's features to output one score. The $R$ ($R=10$, $25$ or $100$) top and bottom scores are then concatenated and fed into a MLP with 128 and 64 hidden neurons and sigmoid activations. 

\paragraph{DeepMIL} We reimplemented the attention based model proposed by \cite{ilse2018attention}. A linear layer with $N$ neurons ($N=64$, $128$ or $256$ here) is applied to the embedding followed by a Gated Attention layer with $N$ hidden neurons. A linear layer followed by a sigmoid activation is then applied to the output.

For both Chowder and DeepMIL, a random subset of $n=8000$ tiles per WSI was used to accelerate training. All hyperparameters were tuned on the different training sets (see Results). Chowder and DeepMIL were trained with a learning rate of 0.001 using cross-entropy loss and the Adam optimizer [\cite{kingma2014adam}]. When multiple WSI were available for a given patient, we averaged the predictions of the models at test time.

\subsection{Datasets}

\begin{table}[t!]
\centering
\sisetup{
    group-separator={,},
    group-minimum-digits=3,
    table-number-alignment = right,
    table-figures-integer = 6,
    detect-weight = true,
    detect-inline-weight = math
}
\resizebox{\columnwidth}{!}{%
\begin{tabular}{lcccc}
\toprule
    & Size & MSI positive & Location & Origin \\
\toprule
 TCGA-CRC & 555 & 78 (14\%) &Colon (74\%) - Rectum (26\%) &  US, 36 centers \\
 TCGA-CRC-Kather  & 360 & 65 (18\%) &Colon (74\%)- Rectum (26\%) & US, 34 centers \\
 TCGA-Gastric  & 375 & 64 (17\%) &Gastric &  US, 22 centers\\
 TCGA-Gastric-Kather  & 284 & 60 (21\%) &Gastric & US, 20 centers\\
 PAIP  & 47 & 12 (26 \%) & Colon &  Korea, 3 centers\\
\bottomrule
\end{tabular}
}
\caption{Datasets used in this study.}
\label{table:datasets}
\end{table}

Three different cohorts were used in this study and are summarized in Table \ref{table:datasets}. For all cohorts, only FFPE images were used. 

TCGA-CRC is a dataset of $n=555$ patients from 36 centers in the US with colorectal tumors. It is a combination of two cohorts from TCGA: TCGA-COAD (colon adenocarcinoma) and TCGA-READ (rectum adenocarcinoma). TCGA-STAD (stomach adenocarcinoma), later referred as TCGA-Gastric, is a dataset of $n=375$ patients from 22 centers in the US with gastric cancer. For both datasets, MSS/MSI-H labels defined by PCR assays were retrieved using TCGA-biolinks [\cite{colaprico2016tcgabiolinks}]. As recommended by ESMO guidelines [\cite{luchini2019esmo}],  MSI-L patients were classified as MSS. TCGA-CRC-Kather and TCGA-Gastric-Kather are variants of respectively TCGA-CRC and TCGA-Gastric datasets published by \cite{kather2019deep}. They were used here for comparison purposes. These datasets consist of a lower number of cases because MSI-L patients were excluded. The exact same MSI labels were used. 

PAIP (cohort from the Pathology AI Platform, http://www.wisepaip.org) is a dataset of $n=47$ patients from 3 centers in Korea with colorectal tumors. MSS/MSI-H labels were determined using PCR assays.

\section{Results}

\begin{table}
\center{
\sisetup{
    group-separator={,},
    group-minimum-digits=3,
    table-number-alignment = right,
    table-figures-integer = 6,
    detect-weight = true,
    detect-inline-weight = math
}

\begin{tabular}{lcc}
\toprule
   Methods &\textsc{TCGA-CRC-Kather} &\textsc{TCGA-Gastric-Kather}\\
\hline 
    \small{\textit{\cite{kather2019deep}}} & 0.77 (0.62-0.87) &0.81 (0.69 - 0.90)  \\
    \small{\textit{\cite{yamashita2021deep}}} & 0.82  (0.71-0.91) & - \\
         \small{\textit{\cite{bilal2021novel}}} & 0.90 & - \\
         \hline
          Ours - MeanPool & 0.85  (0.76 - 0.94) & 0.78  (0.68 - 0.88) \\
         \textbf{Ours - Chowder} & \textbf{0.92 (0.84 - 0.99)} & \textbf{0.83 (0.75 - 0.92)} \\
         Ours - DeepMIL & 0.85 (0.75 - 0.94) & 0.79 (0.69 - 0.89) \\

\bottomrule
\end{tabular}
}
\caption{AUCs on \cite{kather2019deep} train/test split. 95\% CI are computed following [\cite{delong1988comparing}] for our models, and using boostrapping for \cite{kather2019deep} and \cite{yamashita2021deep}.}
\label{table:KatherSplitResult}
\end{table}

\subsection{Cross-validations on TCGA-CRC and TCGA-Gastric}
We first compared three multiple instance learning models, MeanPool, Chowder and DeepMIL, using the MoCo-CRC features on the TCGA-CRC-Kather cohort, and the MoCo-Gastric features on the TCGA-Gastric-Kather cohort. For the sake of comparison, we used the exact same train / test split as in [\cite{kather2019deep}] and compared our results with the ones published in the literature on this split.

We tuned the hyperparameters of Chowder ($R=10$, $25$ or $100$), DeepMIL (size of attention layer = 64, 128 or 256), MeanPool (l2 penalization = 0, 0.5 or 1) and the number of epochs (ranging from 5 to 120) using a grid search on the training sets. For both TCGA-CRC-Kather and TCGA-Gastric-Kather, Chowder model performed best and obtained respectively AUCs of 0.92 and 0.83, achieving state-of-the-art results on these datasets (see Table \ref{table:KatherSplitResult}).

To analyze the gain of MoCo features over ImageNet ones, we ran larger cross-validation experiments using 15 distinct splits on the full TCGA cohorts (5 fold cross-validation, repeated 3 times) and reported results in Table \ref{table:CV}. For a fair comparison, we tuned the hyperparameters of all models on the training set of the Kather split. In all our experiments, MoCo substantially outperformed ImageNet in both TCGA-CRC and TCGA-Gastric cohorts. Results obtained with MoCo features were also better than the ones previously reported with cross-validation. We also report cross-validation results using center split in Supplementary Table \ref{table:expCVSup}.

\begin{table}
\center{
\sisetup{
    group-separator={,},
    group-minimum-digits=3,
    table-number-alignment = right,
    table-figures-integer = 6,
    detect-weight = true,
    detect-inline-weight = math
}

\begin{tabular}{lcccc}
\toprule
    &\multicolumn{2}{c}{\textsc{TCGA-CRC}} &\multicolumn{2}{c}{\textsc{TCGA-Gastric}}\\
    \hline
             \cite{echle2020clinical} &\multicolumn{2}{c}{ 0.74 (0.66–0.80)} &\multicolumn{2}{c}{ - }\\
         \cite{kather2020pan} &\multicolumn{2}{c}{- } &\multicolumn{2}{c}{ 0.72 }\\
         \cite{bilal2021novel} &\multicolumn{2}{c}{0.86}&\multicolumn{2}{c}{ - }\\
    
\toprule
    & ImageNet & MoCo-CRC & ImageNet & MoCo-Gastric\\
\cmidrule(r){2-3}\cmidrule(l){4-5}
Ours - MeanPool & 0.84 (0.05) & 0.87 (0.05) \textcolor{greend}{\small{+0.03}} & 0.76 (0.04) & 0.82 (0.05) \textcolor{greend}{\small{+0.06}} \\
    Ours - Chowder  & 0.81 (0.05) & 0.88 (0.04) \textcolor{greend}{\small{+0.07}} & 0.73 (0.07) & 0.83 (0.06)
    \textcolor{greend}{\small{+0.11}}  \\
Ours - DeepMIL  & 0.82 (0.05) & 0.88 (0.05) \textcolor{greend}{\small{+0.06}} & 0.74 (0.01) & 0.85 (0.05) \textcolor{greend}{\small{+0.11}} \\


\bottomrule
\end{tabular}
}
\caption{Performances for $3 \times 5$ folds cross-validation (AUC), means, and standard deviations on TCGA-CRC and TCGA-Gastric datasets. Mean, lower and upper bounds on $3$ folds are reported by \cite{echle2020clinical}, means on respectively $4$ folds and $3$ folds are reported by \cite{bilal2021novel} and \cite{kather2020pan}}
\label{table:CV}
\end{table}

\subsection{External validation on CRC dataset PAIP}

\begin{table}[!ht]
\center{
\sisetup{
    group-separator={,},
    group-minimum-digits=3,
    table-number-alignment = right,
    table-figures-integer = 6,
    detect-weight = true,
    detect-inline-weight = math
}

\begin{tabular}{lcc}
\toprule
   Methods &\textsc{ImageNet} &\textsc{MoCo-CRC}\\
\hline 
    MeanPool  & 0.61 (0.42 - 0.80) & 0.82 (0.66 - 0.99) \textcolor{greend}{\small{+0.21}} \\
    Chowder & 0.86 (0.74 - 0.97) & 0.97 (0.93 - 1.00) \textcolor{greend}{\small{+0.11}} \\
    DeepMIL & 0.83 (0.67 - 1.0) & 0.90 (0.78 - 1.00) \textcolor{greend}{\small{+0.07}} \\
    
\bottomrule
\end{tabular}
}
\caption{AUCs on PAIP for models trained on TCGA-CRC. \cite{bilal2021novel} report AUC of 0.98 but no CI interval is given.}
\label{table:PAIP}
\end{table}

A limitation of the previous experiments is that ResNet using SSL were pretrained on the full TCGA cohorts, slightly breaking the train/test split independence assumption even if no MSI/MSS labels were used. Thus, we further evaluated MSI detection on an independent cohort of $n=47$ colon cases from PAIP organisation. We used the median prediction of the ensemble of models trained during cross-validation in the previous experiment. Chowder with MoCo-CRC features yielded an AUC of 0.97 on par with the 0.98 reported in \cite{bilal2021novel}, and significantly higher than Chowder with ImageNet features (AUC of 0.82, $p=0.03$ with DeLong's test [\cite{delong1988comparing}]). The superiority of MoCo-CRC features was observed for all models (Table \ref{table:PAIP}), confirming that the MoCo-CRC pretrained backbone is a more robust feature extractor than the ImageNet for histology images.

\subsection{Transfer CRC to Gastric}

We assessed the performances for MSI detection when transferring models trained on TCGA-CRC to TCGA-Gastric with three different feature extractors: ImageNet, MoCo-CRC, MoCo-CRC-Gastric. For fair comparison, all hyperparameters were tuned on the training split of TCGA-CRC-Kather (Supplementary Table \ref{table:HyperParameters}).

MoCo-CRC-Gastric models consistently yielded the best results on TCGA-Gastric with AUCs up to 0.80 (Table \ref{table:transfer}) while also demonstrating high performances on TCGA-CRC (Supplementary Table \ref{table:CV_MOCO_CRC_GASTRIC}). Surprisingly, the ImageNet MeanPool model, which performed poorly on PAIP, also obtained a high AUC of 0.76, however significantly lower compared to 0.80 ($p=0.05$ with DeLong's test). To our knowledge, this is the best performance reported in a transfer setting from one organ to another.

\begin{table}
\center{
\sisetup{
    group-separator={,},
    group-minimum-digits=3,
    table-number-alignment = right,
    table-figures-integer = 6,
    detect-weight = true,
    detect-inline-weight = math
}
\begin{tabular}{lccc}
\toprule
    Methods & ImageNet & MoCo-CRC & MoCo-CRC-Gastric\\
\hline 
MeanPool &0.76 (0.69-0.82)&0.67 (0.59-0.74)&0.76 (0.69-0.82) \\
Chowder  &0.71 (0.64-0.78)&0.73 (0.66-0.80)& 0.78 (0.73-0.84) \\
DeepMIL  &0.72 (0.66-0.78)&0.71 (0.64-0.79)& 0.80 (0.75-0.86) \\

\bottomrule
\end{tabular}
}
\caption{AUCs for models trained on TCGA-CRC and evaluated on TCGA-Gastric with different feature extractors.}
\label{table:transfer}
\end{table}

\section{Interpretability}
\begin{figure}[!ht]
\centering
\includegraphics[width=12cm]{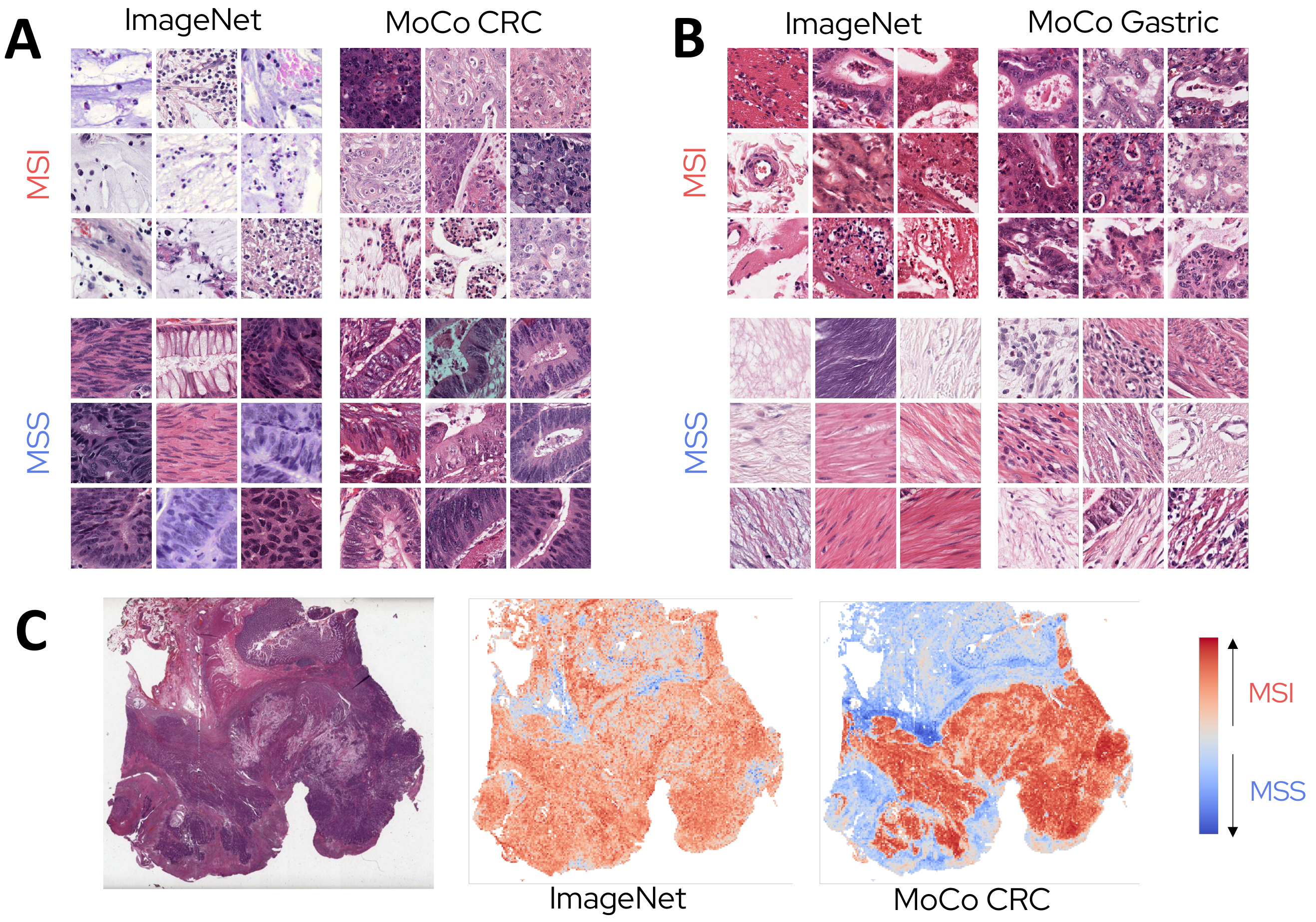}
\caption{Most predictive regions identified by Chowder
A. Tiles most predictive of MSI and MSS phenotypes in CRC. B. Similar for gastric cancer. C. Heatmaps obtained with ImageNet or MoCo-CRC features.}
\label{fig:interpretability}
\end{figure}

Within Chowder, each tile is associated with a single score and a MLP is applied to the top and lowest scores of the WSI (Section 3.1). To explore potential differences in interpretability when using ImageNet or MoCo features, a pathologist expert in dMMR tumors reviewed the subset of tiles associated with extreme scores within TCGA-CRC and TCGA-Gastric cohorts, for both feature extractors. 

Among the top scored tiles (Figure \ref{fig:interpretability} A, B), the pathologist recognized histological patterns which have been previously described as associated with MSI tumors [\cite{greenson2009pathologic}], including tumor lymphocyte infiltration, mucinous differentiation and presence of dirty necrosis - the latter being more frequently detected in gastric tumors. While ImageNet-based models outputs were based on patterns within and outside tumor regions (such as lymphocyte infiltration in the tumor microenvironment), MoCo-based models clearly focused on tumor epithelium, highlighting both poor differentiation of epithelial cells and tumor infiltrating lymphocytes (TILs) as patterns associated with MSI tumors. Specifically, MoCo models better segmented tumoral regions, as shown by the tile scores heatmap at the WSI level (Figure  \ref{fig:interpretability} C). 

Overall, lowest scored tiles were unrelated to known MSI patterns, displaying non tumoral tissues such as muscle cells. In the TCGA-CRC cohort, the MoCo-CRC model additionally highlighted the presence of intratumoral epithelial cells without MSI tumor characteristics, showing once again a higher focus on tumor regions, in line with the regions actually inspected by a pathologist when searching for MSI related patterns.

\section{Discussion}

In this work, we showed that feature extractors pretrained using SSL (MoCo V2) on TCGA reach state-of-the-art results for MSI prediction both in colorectal and gastric cancers (Table \ref{table:KatherSplitResult}). Extensive cross-validations showed the clear superiority of these models over their counterparts pretrained using ImageNet (Table \ref{table:CV}). In addition, the former generalize better on an external CRC cohort (Table \ref{table:PAIP}). Finally, we observed that using a feature extractor pretrained on several organs using SSL (Table \ref{table:transfer}) opens the way to both state-of-the-art performances in cross-validation and robust generalization from one organ to another.

There are several limitations to our current study. First, our models could benefit from being trained on more data as shown in the training curve from \cite{echle2020clinical} (Figure 1.c). Second, our models should be validated on larger cohorts, encompassing different patient populations, treatments (neoadjuvant chemotherapy can impact cell morphology), scanner manufacturer and sample types (resections, biopsies). Third, SSL techniques are evolving rapidly using new architectures such as Vision Transformers [\cite{dosovitskiy2020image}] and more experiments are required to find how to apply them in histology, including taking into account the spatial arrangement of the tiles. However, such experiments require access to extensive computational resources, which limits reproducibility. Finally, in contrast to several concurrent works [\cite{echle2020clinical, bilal2021novel, kather2019deep}] that fine-tuned the backbones, while our study kept them entirely frozen.

A recent study, \cite{kacew2021artificial} showed that performances of deep learning models may impact the diagnosis of MSI for patients with CRC. Notably, MSI diagnosis is not routinely done for patients with other solid tumors, missing the identification of candidate patients for immunotherapy. Our results indicate that SSL is a promising solution to develop accurate models for frequent tumor localization such as  oesophagus, pancreas, small intestine or even brain, where images are available but MSI prevalence is too low for systematic IHC or molecular testing.


\newpage

\acks{

We thank Florence Renaud, pathologist at CHU Lille, for her insightful analysis of the most predictive regions. We thank Patrick Sin-Chan for his corrections of the manuscript.

This work was granted access to the HPC resources of IDRIS under the allocation AD011012519
made by GENCI.

The results published here are part based upon data generated by the TCGA Research Network: https://www.cancer.gov/tcga.

Regarding the PAIP dataset: De-identified pathology images and annotations used in this research were prepared and provided by the Seoul National University Hospital by a grant of the Korea Health Technology R\&D Project through the Korea Health Industry Development Institute (KHIDI), funded by the Ministry of Health \& Welfare, Republic of Korea (grant number: HI18C0316).
}

\appendix
\beginsupplement
\section*{Supplementary Tables and Figures}

\begin{table}[H]
\resizebox{\columnwidth}{!}{%
\begin{tabular}{lccccc}
\toprule
    &&\multicolumn{2}{c}{\textsc{TCGA-CRC}} &\multicolumn{2}{c}{\textsc{TCGA-Gastric}}\\
  \cmidrule(lr){2-2}\cmidrule(lr){3-4}\cmidrule(l){5-6}
    &Split & ImageNet & MoCo-CRC & ImageNet & MoCo-Gastric\\
\cmidrule(lr){2-2}\cmidrule(lr){3-4}\cmidrule(l){5-6}

MeanPool &CV & 0.84 (0.05) & 0.87 (0.05) \textcolor{greend}{\small{+0.03}} & 0.76 (0.04) & 0.82 (0.05) \textcolor{greend}{\small{+0.06}} \\

MeanPool &CV centers&  0.78 (0.10) & 0.85 (0.07) \textcolor{greend}{\small{+0.07}}& 0.72 (0.12) & 0.85 (0.07) \textcolor{greend}{\small{+0.13}} \\

DeepMIL  &CV& 0.82 (0.05) & 0.88 (0.05) \textcolor{greend}{\small{+0.06}} & 0.74 (0.01) & 0.85 (0.05) \textcolor{greend}{\small{+0.11}} \\

DeepMIL  &CV centers& 0.79 (0.059) & 0.84 (0.11) \textcolor{greend}{\small{+0.05}} &0.73 (0.15) & 0.85 (0.05) \textcolor{greend}{\small{+0.12}} \\

    Chowder  &CV& 0.81 (0.05) & 0.88 (0.04) \textcolor{greend}{\small{+0.07}} & 0.73 (0.07) & 0.84 (0.06) \textcolor{greend}{\small{+0.11}}  \\
    
    Chowder  &CV centers&0.75 (0.15) & 0.83 (0.12) \textcolor{greend}{\small{+0.08}} &0.72 (0.08) & 0.86 (0.05) \textcolor{greend}{\small{+0.14}} \\
\bottomrule
\end{tabular}
}

\caption{Cross-validation performances (AUC) on TCGA-CRC and TCGA-Gastric. We report mean and standard deviation on $3 \times 5$ folds. We split the data into 5 fold either randomly (CV) or by ensuring that all samples from a center are in the same set (CV centers).
}
\label{table:expCVSup}
\end{table}

\begin{table}[!ht]
\center{
\sisetup{
    group-separator={,},
    group-minimum-digits=3,
    table-number-alignment = right,
    table-figures-integer = 6,
    detect-weight = true,
    detect-inline-weight = math
}
\begin{tabular}{lcccc}
\toprule

    Feature extractor &Train dataset & MeanPool & Chowder &DeepMIL\\
\hline 
ImageNet & CRC & C = 1  & R = 100 & N = 64  \\
&&& 100 epochs &30 epochs \\
\hline
ImageNet & Gastric & C = 0.0  & R = 25 & N = 64 \\
&&& 30 epochs &20 epochs \\
\hline
MoCo-CRC & CRC  & C = 0.5  & R = 10 & N = 32 \\
&&& 10 epochs &10 epochs \\
\hline
MoCo-Gastric & Gastric & C = 1.0  & R = 100 & N = 128 \\
&&& 30 epochs &10 epochs \\
\hline
MoCo-CRC-Gastric & CRC  & C = 0.5  & R = 100 & N = 64 \\
&&& 30 epochs &10 epochs \\
\hline

\end{tabular}
}
\caption{Hyperparameters used for the different models. All hyperparameters were tuned on the training sets of the TCGA-CRC-Kather or TCGA-Gastric-Kather cohorts. $C$ refers to L2 penalization coefficient, $R$ to the number of extreme tiles used in Chowder and $N$ to the size of the attention layer in DeepMIL.}
\label{table:HyperParameters}
\end{table}

\begin{table}[!ht]
\center{
\sisetup{
    group-separator={,},
    group-minimum-digits=3,
    table-number-alignment = right,
    table-figures-integer = 6,
    detect-weight = true,
    detect-inline-weight = math
}
\begin{tabular}{lc}
\toprule

    Methods & MoCo-CRC-Gastric\\
\hline 
MeanPool &0.86 (0.04) \\
Chowder  &0.88 (0.05) \\
DeepMIL  &0.88 (0.06) \\

\bottomrule
\end{tabular}
}
\caption{Cross-validation performances (AUC) on TCGA-CRC, for the models trained using the features of MoCo-CRC-Gastric. The result reported are the mean and the standard deviation from 5-fold cross-validation repeated 3 times.}
\label{table:CV_MOCO_CRC_GASTRIC}
\end{table}

\begin{figure}[H]
\centering
\includegraphics[width=12cm]{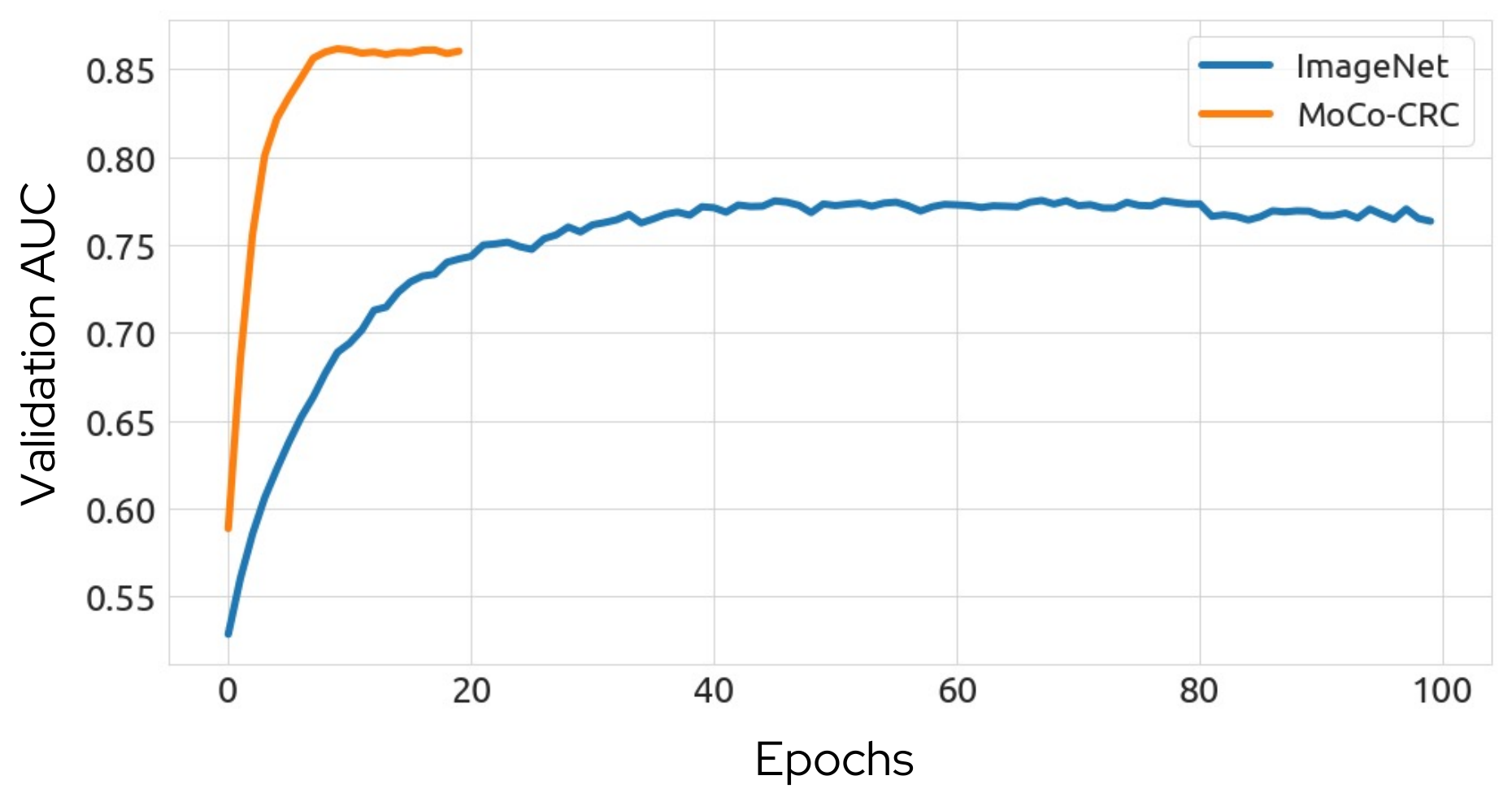}
\caption{Validation Curves of Chowder with ImageNet and MoCo-CRC features. The plot represents the average validation AUCs per epoch of all runs of the cross-validation for Chowder trained with ImageNet and MoCo-CRC features, depicted in Table 3.
}
\label{fig:architecture}
\end{figure}

\newpage
\bibliography{bibli}

\begin{thebibliography}{45}
\providecommand{\natexlab}[1]{#1}
\providecommand{\url}[1]{\texttt{#1}}
\expandafter\ifx\csname urlstyle\endcsname\relax
  \providecommand{\doi}[1]{doi: #1}\else
  \providecommand{\doi}{doi: \begingroup \urlstyle{rm}\Url}\fi

\bibitem[Abbet et~al.(2021)Abbet, Studer, Fischer, Dawson, Zlobec, Bozorgtabar,
  and Thiran]{abbet2021self}
Christian Abbet, Linda Studer, Andreas Fischer, Heather Dawson, Inti Zlobec,
  Behzad Bozorgtabar, and Jean-Philippe Thiran.
\newblock Self-rule to adapt: Learning generalized features from
  sparsely-labeled data using unsupervised domain adaptation for colorectal
  cancer tissue phenotyping.
\newblock In \emph{Medical Imaging with Deep Learning}, 2021.

\bibitem[Bilal et~al.(2021)Bilal, Raza, Azam, Graham, Ilyas, Cree, Snead,
  Minhas, and Rajpoot]{bilal2021novel}
Mohsin Bilal, Shan E~Ahmed Raza, Ayesha Azam, Simon Graham, Muhammad Ilyas,
  Ian~A Cree, David Snead, Fayyaz Minhas, and Nasir~M Rajpoot.
\newblock Novel deep learning algorithm predicts the status of molecular
  pathways and key mutations in colorectal cancer from routine histology
  images.
\newblock \emph{medRxiv}, 2021.

\bibitem[Cao et~al.(2020)Cao, Yang, Ma, Liu, Zhao, Li, Wu, Wang, Lu, Cai,
  et~al.]{cao2020development}
Rui Cao, Fan Yang, Si-Cong Ma, Li~Liu, Yu~Zhao, Yan Li, De-Hua Wu, Tongxin
  Wang, Wei-Jia Lu, Wei-Jing Cai, et~al.
\newblock Development and interpretation of a pathomics-based model for the
  prediction of microsatellite instability in colorectal cancer.
\newblock \emph{Theranostics}, 10\penalty0 (24):\penalty0 11080, 2020.

\bibitem[Caron et~al.(2021)Caron, Touvron, Misra, J{\'e}gou, Mairal,
  Bojanowski, and Joulin]{caron2021emerging}
Mathilde Caron, Hugo Touvron, Ishan Misra, Herv{\'e} J{\'e}gou, Julien Mairal,
  Piotr Bojanowski, and Armand Joulin.
\newblock Emerging properties in self-supervised vision transformers.
\newblock \emph{arXiv preprint arXiv:2104.14294}, 2021.

\bibitem[Chen et~al.(2020{\natexlab{a}})Chen, Kornblith, Norouzi, and
  Hinton]{chen2020simple}
Ting Chen, Simon Kornblith, Mohammad Norouzi, and Geoffrey Hinton.
\newblock A simple framework for contrastive learning of visual
  representations.
\newblock In \emph{International conference on machine learning}, pages
  1597--1607. PMLR, 2020{\natexlab{a}}.

\bibitem[Chen et~al.(2020{\natexlab{b}})Chen, Kornblith, Swersky, Norouzi, and
  Hinton]{chen2020big}
Ting Chen, Simon Kornblith, Kevin Swersky, Mohammad Norouzi, and Geoffrey
  Hinton.
\newblock Big self-supervised models are strong semi-supervised learners.
\newblock \emph{arXiv preprint arXiv:2006.10029}, 2020{\natexlab{b}}.

\bibitem[Chen et~al.(2020{\natexlab{c}})Chen, Fan, Girshick, and
  He]{chen2020improved}
Xinlei Chen, Haoqi Fan, Ross Girshick, and Kaiming He.
\newblock Improved baselines with momentum contrastive learning.
\newblock \emph{arXiv preprint arXiv:2003.04297}, 2020{\natexlab{c}}.

\bibitem[Chen et~al.(2021)Chen, Xie, and He]{chen2021empirical}
Xinlei Chen, Saining Xie, and Kaiming He.
\newblock An empirical study of training self-supervised vision transformers.
\newblock \emph{arXiv preprint arXiv:2104.02057}, 2021.

\bibitem[Colaprico et~al.(2016)Colaprico, Silva, Olsen, Garofano, Cava,
  Garolini, Sabedot, Malta, Pagnotta, Castiglioni,
  et~al.]{colaprico2016tcgabiolinks}
Antonio Colaprico, Tiago~C Silva, Catharina Olsen, Luciano Garofano, Claudia
  Cava, Davide Garolini, Thais~S Sabedot, Tathiane~M Malta, Stefano~M Pagnotta,
  Isabella Castiglioni, et~al.
\newblock Tcgabiolinks: an r/bioconductor package for integrative analysis of
  tcga data.
\newblock \emph{Nucleic acids research}, 44\penalty0 (8):\penalty0 e71--e71,
  2016.

\bibitem[Courtiol et~al.(2018)Courtiol, Tramel, Sanselme, and
  Wainrib]{courtiol2018classification}
Pierre Courtiol, Eric~W Tramel, Marc Sanselme, and Gilles Wainrib.
\newblock Classification and disease localization in histopathology using only
  global labels: A weakly-supervised approach.
\newblock \emph{arXiv preprint arXiv:1802.02212}, 2018.

\bibitem[Dehaene et~al.(2020)Dehaene, Camara, Moindrot, de~Lavergne, and
  Courtiol]{dehaene2020self}
Olivier Dehaene, Axel Camara, Olivier Moindrot, Axel de~Lavergne, and Pierre
  Courtiol.
\newblock Self-supervision closes the gap between weak and strong supervision
  in histology.
\newblock \emph{arXiv preprint arXiv:2012.03583}, 2020.

\bibitem[DeLong et~al.(1988)DeLong, DeLong, and
  Clarke-Pearson]{delong1988comparing}
Elizabeth~R DeLong, David~M DeLong, and Daniel~L Clarke-Pearson.
\newblock Comparing the areas under two or more correlated receiver operating
  characteristic curves: a nonparametric approach.
\newblock \emph{Biometrics}, pages 837--845, 1988.

\bibitem[Deng et~al.(2009)Deng, Dong, Socher, Li, Li, and
  Fei-Fei]{deng2009imagenet}
Jia Deng, Wei Dong, Richard Socher, Li-Jia Li, Kai Li, and Li~Fei-Fei.
\newblock Imagenet: A large-scale hierarchical image database.
\newblock In \emph{2009 IEEE conference on computer vision and pattern
  recognition}, pages 248--255. Ieee, 2009.

\bibitem[Dosovitskiy et~al.(2020)Dosovitskiy, Beyer, Kolesnikov, Weissenborn,
  Zhai, Unterthiner, Dehghani, Minderer, Heigold, Gelly,
  et~al.]{dosovitskiy2020image}
Alexey Dosovitskiy, Lucas Beyer, Alexander Kolesnikov, Dirk Weissenborn,
  Xiaohua Zhai, Thomas Unterthiner, Mostafa Dehghani, Matthias Minderer, Georg
  Heigold, Sylvain Gelly, et~al.
\newblock An image is worth 16x16 words: Transformers for image recognition at
  scale.
\newblock \emph{arXiv preprint arXiv:2010.11929}, 2020.

\bibitem[Echle et~al.(2020)Echle, Grabsch, Quirke, van~den Brandt, West,
  Hutchins, Heij, Tan, Richman, Krause, et~al.]{echle2020clinical}
Amelie Echle, Heike~Irmgard Grabsch, Philip Quirke, Piet~A van~den Brandt,
  Nicholas~P West, Gordon~GA Hutchins, Lara~R Heij, Xiuxiang Tan, Susan~D
  Richman, Jeremias Krause, et~al.
\newblock Clinical-grade detection of microsatellite instability in colorectal
  tumors by deep learning.
\newblock \emph{Gastroenterology}, 159\penalty0 (4):\penalty0 1406--1416, 2020.

\bibitem[Fujiyoshi et~al.(2017)Fujiyoshi, Yamaguchi, Kakuta, Takahashi, Arai,
  Yamada, Yamamoto, Ohde, Takao, Horiguchi, et~al.]{fujiyoshi2017predictive}
Kenji Fujiyoshi, Tatsuro Yamaguchi, Miho Kakuta, Akemi Takahashi, Yoshiko Arai,
  Mina Yamada, Gou Yamamoto, Sachiko Ohde, Misato Takao, Shin-ichiro Horiguchi,
  et~al.
\newblock Predictive model for high-frequency microsatellite instability in
  colorectal cancer patients over 50 years of age.
\newblock \emph{Cancer medicine}, 6\penalty0 (6):\penalty0 1255--1263, 2017.

\bibitem[Gildenblat and Klaiman(2019)]{gildenblat2019self}
Jacob Gildenblat and Eldad Klaiman.
\newblock Self-supervised similarity learning for digital pathology.
\newblock \emph{arXiv preprint arXiv:1905.08139}, 2019.

\bibitem[Greenson et~al.(2009)Greenson, Huang, Herron, Moreno, Bonner, Tomsho,
  Ben-Izhak, Cohen, Trougouboff, Bejhar, et~al.]{greenson2009pathologic}
Joel~K Greenson, Shu-Chen Huang, Casey Herron, Victor Moreno, Joseph~D Bonner,
  Lynn~P Tomsho, Ofer Ben-Izhak, Hector~I Cohen, Phillip Trougouboff, Jacob
  Bejhar, et~al.
\newblock Pathologic predictors of microsatellite instability in colorectal
  cancer.
\newblock \emph{The American journal of surgical pathology}, 33\penalty0
  (1):\penalty0 126, 2009.

\bibitem[Grill et~al.(2020)Grill, Strub, Altch{\'e}, Tallec, Richemond,
  Buchatskaya, Doersch, Pires, Guo, Azar, et~al.]{grill2020bootstrap}
Jean-Bastien Grill, Florian Strub, Florent Altch{\'e}, Corentin Tallec,
  Pierre~H Richemond, Elena Buchatskaya, Carl Doersch, Bernardo~Avila Pires,
  Zhaohan~Daniel Guo, Mohammad~Gheshlaghi Azar, et~al.
\newblock Bootstrap your own latent: A new approach to self-supervised
  learning.
\newblock \emph{arXiv preprint arXiv:2006.07733}, 2020.

\bibitem[He et~al.(2016)He, Zhang, Ren, and Sun]{he2016deep}
Kaiming He, Xiangyu Zhang, Shaoqing Ren, and Jian Sun.
\newblock Deep residual learning for image recognition.
\newblock In \emph{Proceedings of the IEEE conference on computer vision and
  pattern recognition}, pages 770--778, 2016.

\bibitem[He et~al.(2020)He, Fan, Wu, Xie, and Girshick]{he2020momentum}
Kaiming He, Haoqi Fan, Yuxin Wu, Saining Xie, and Ross Girshick.
\newblock Momentum contrast for unsupervised visual representation learning.
\newblock In \emph{Proceedings of the IEEE/CVF Conference on Computer Vision
  and Pattern Recognition}, pages 9729--9738, 2020.

\bibitem[Hildebrand et~al.(2021)Hildebrand, Pierce, Dennis, Paracha, and
  Maoz]{hildebrand2021artificial}
Lindsey~A Hildebrand, Colin~J Pierce, Michael Dennis, Munizay Paracha, and Asaf
  Maoz.
\newblock Artificial intelligence for histology-based detection of
  microsatellite instability and prediction of response to immunotherapy in
  colorectal cancer.
\newblock \emph{Cancers}, 13\penalty0 (3):\penalty0 391, 2021.

\bibitem[Hong et~al.(2020)Hong, Liu, DeLair, Razavian, and
  Feny{\"o}]{hong2020predicting}
Runyu Hong, Wenke Liu, Deborah DeLair, Narges Razavian, and David Feny{\"o}.
\newblock Predicting endometrial cancer subtypes and molecular features from
  histopathology images using multi-resolution deep learning models.
\newblock \emph{bioRxiv}, 2020.

\bibitem[Hyde et~al.(2010)Hyde, Fontaine, Stuckless, Green, Pollett, Simms,
  Sipahimalani, Parfrey, and Younghusband]{hyde2010histology}
Angela Hyde, Daniel Fontaine, Susan Stuckless, Roger Green, Aaron Pollett,
  Michelle Simms, Payal Sipahimalani, Patrick Parfrey, and Banfield
  Younghusband.
\newblock A histology-based model for predicting microsatellite instability in
  colorectal cancers.
\newblock \emph{The American journal of surgical pathology}, 34\penalty0
  (12):\penalty0 1820--1829, 2010.

\bibitem[Ilse et~al.(2018)Ilse, Tomczak, and Welling]{ilse2018attention}
Maximilian Ilse, Jakub Tomczak, and Max Welling.
\newblock Attention-based deep multiple instance learning.
\newblock In \emph{International conference on machine learning}, pages
  2127--2136. PMLR, 2018.

\bibitem[Jenkins et~al.(2007)Jenkins, Hayashi, O’shea, Burgart, Smyrk,
  Shimizu, Waring, Ruszkiewicz, Pollett, Redston, et~al.]{jenkins2007pathology}
Mark~A Jenkins, Shinichi Hayashi, Anne-Marie O’shea, Lawrence~J Burgart,
  Tom~C Smyrk, David Shimizu, Paul~M Waring, Andrew~R Ruszkiewicz, Aaron~F
  Pollett, Mark Redston, et~al.
\newblock Pathology features in bethesda guidelines predict colorectal cancer
  microsatellite instability: a population-based study.
\newblock \emph{Gastroenterology}, 133\penalty0 (1):\penalty0 48--56, 2007.

\bibitem[Kacew et~al.(2021)Kacew, Strohbehn, Saulsberry, Laiteerapong,
  Cipriani, Kather, and Pearson]{kacew2021artificial}
Alec~J Kacew, Garth~W Strohbehn, Loren Saulsberry, Neda Laiteerapong, Nicole~A
  Cipriani, Jakob~N Kather, and Alexander~T Pearson.
\newblock Artificial intelligence can cut costs while maintaining accuracy in
  colorectal cancer genotyping.
\newblock \emph{Frontiers in Oncology}, 11, 2021.

\bibitem[Kather et~al.(2019)Kather, Pearson, Halama, J{\"a}ger, Krause, Loosen,
  Marx, Boor, Tacke, Neumann, et~al.]{kather2019deep}
Jakob~Nikolas Kather, Alexander~T Pearson, Niels Halama, Dirk J{\"a}ger,
  Jeremias Krause, Sven~H Loosen, Alexander Marx, Peter Boor, Frank Tacke,
  Ulf~Peter Neumann, et~al.
\newblock Deep learning can predict microsatellite instability directly from
  histology in gastrointestinal cancer.
\newblock \emph{Nature medicine}, 25\penalty0 (7):\penalty0 1054--1056, 2019.

\bibitem[Kather et~al.(2020)Kather, Heij, Grabsch, Loeffler, Echle, Muti,
  Krause, Niehues, Sommer, Bankhead, et~al.]{kather2020pan}
Jakob~Nikolas Kather, Lara~R Heij, Heike~I Grabsch, Chiara Loeffler, Amelie
  Echle, Hannah~Sophie Muti, Jeremias Krause, Jan~M Niehues, Kai~AJ Sommer,
  Peter Bankhead, et~al.
\newblock Pan-cancer image-based detection of clinically actionable genetic
  alterations.
\newblock \emph{Nature Cancer}, 1\penalty0 (8):\penalty0 789--799, 2020.

\bibitem[Kingma and Ba(2014)]{kingma2014adam}
Diederik~P Kingma and Jimmy Ba.
\newblock Adam: A method for stochastic optimization.
\newblock \emph{arXiv preprint arXiv:1412.6980}, 2014.

\bibitem[Koohbanani et~al.(2021)Koohbanani, Unnikrishnan, Khurram,
  Krishnaswamy, and Rajpoot]{koohbanani2021self}
Navid~Alemi Koohbanani, Balagopal Unnikrishnan, Syed~Ali Khurram, Pavitra
  Krishnaswamy, and Nasir Rajpoot.
\newblock Self-path: Self-supervision for classification of pathology images
  with limited annotations.
\newblock \emph{IEEE Transactions on Medical Imaging}, 2021.

\bibitem[Le et~al.(2017)Le, Durham, Smith, Wang, Bartlett, Aulakh, Lu,
  Kemberling, Wilt, Luber, et~al.]{le2017mismatch}
Dung~T Le, Jennifer~N Durham, Kellie~N Smith, Hao Wang, Bjarne~R Bartlett,
  Laveet~K Aulakh, Steve Lu, Holly Kemberling, Cara Wilt, Brandon~S Luber,
  et~al.
\newblock Mismatch repair deficiency predicts response of solid tumors to pd-1
  blockade.
\newblock \emph{Science}, 357\penalty0 (6349):\penalty0 409--413, 2017.

\bibitem[Lee et~al.(2021)Lee, Song, and Jang]{lee2021feasibility}
Sung~Hak Lee, In~Hye Song, and Hyun-Jong Jang.
\newblock Feasibility of deep learning-based fully automated classification of
  microsatellite instability in tissue slides of colorectal cancer.
\newblock \emph{International Journal of Cancer}, 2021.

\bibitem[Li et~al.(2021{\natexlab{a}})Li, Li, and Eliceiri]{li2021dual}
Bin Li, Yin Li, and Kevin~W Eliceiri.
\newblock Dual-stream multiple instance learning network for whole slide image
  classification with self-supervised contrastive learning.
\newblock In \emph{Proceedings of the IEEE/CVF Conference on Computer Vision
  and Pattern Recognition}, pages 14318--14328, 2021{\natexlab{a}}.

\bibitem[Li et~al.(2021{\natexlab{b}})Li, Yang, Zhang, Gao, Xiao, Dai, Yuan,
  and Gao]{li2021efficient}
Chunyuan Li, Jianwei Yang, Pengchuan Zhang, Mei Gao, Bin Xiao, Xiyang Dai,
  Lu~Yuan, and Jianfeng Gao.
\newblock Efficient self-supervised vision transformers for representation
  learning.
\newblock \emph{arXiv preprint arXiv:2106.09785}, 2021{\natexlab{b}}.

\bibitem[Lu et~al.(2019)Lu, Chen, Wang, Dillon, and Mahmood]{lu2019semi}
Ming~Y Lu, Richard~J Chen, Jingwen Wang, Debora Dillon, and Faisal Mahmood.
\newblock Semi-supervised histology classification using deep multiple instance
  learning and contrastive predictive coding.
\newblock \emph{arXiv preprint arXiv:1910.10825}, 2019.

\bibitem[Luchini et~al.(2019)Luchini, Bibeau, Ligtenberg, Singh, Nottegar,
  Bosse, Miller, Riaz, Douillard, Andre, et~al.]{luchini2019esmo}
C~Luchini, F~Bibeau, MJL Ligtenberg, N~Singh, A~Nottegar, T~Bosse, R~Miller,
  N~Riaz, J-Y Douillard, F~Andre, et~al.
\newblock Esmo recommendations on microsatellite instability testing for
  immunotherapy in cancer, and its relationship with pd-1/pd-l1 expression and
  tumour mutational burden: a systematic review-based approach.
\newblock \emph{Annals of Oncology}, 30\penalty0 (8):\penalty0 1232--1243,
  2019.

\bibitem[Prasad et~al.(2018)Prasad, Kaestner, and Mailankody]{prasad2018cancer}
Vinay Prasad, Victoria Kaestner, and Sham Mailankody.
\newblock Cancer drugs approved based on biomarkers and not tumor type—fda
  approval of pembrolizumab for mismatch repair-deficient solid cancers.
\newblock \emph{JAMA oncology}, 4\penalty0 (2):\penalty0 157--158, 2018.

\bibitem[Rom{\'a}n et~al.(2010)Rom{\'a}n, Verd{\'u}, Calvo, Vidal, Sanjuan,
  Jimeno, Salas, Autonell, Trias, Gonz{\'a}lez,
  et~al.]{roman2010microsatellite}
Ruth Rom{\'a}n, Montse Verd{\'u}, Miquel Calvo, August Vidal, Xavier Sanjuan,
  Mireya Jimeno, Antonio Salas, Josefina Autonell, Isabel Trias, Marta
  Gonz{\'a}lez, et~al.
\newblock Microsatellite instability of the colorectal carcinoma can be
  predicted in the conventional pathologic examination. a prospective
  multicentric study and the statistical analysis of 615 cases consolidate our
  previously proposed logistic regression model.
\newblock \emph{Virchows Archiv}, 456\penalty0 (5):\penalty0 533--541, 2010.

\bibitem[Ronneberger et~al.(2015)Ronneberger, Fischer, and
  Brox]{ronneberger2015u}
Olaf Ronneberger, Philipp Fischer, and Thomas Brox.
\newblock U-net: Convolutional networks for biomedical image segmentation.
\newblock In \emph{International Conference on Medical image computing and
  computer-assisted intervention}, pages 234--241. Springer, 2015.

\bibitem[Sargent et~al.(2010)Sargent, Marsoni, Monges, Thibodeau, Labianca,
  Hamilton, French, Kabat, Foster, Torri, et~al.]{sargent2010defective}
Daniel~J Sargent, Silvia Marsoni, Genevieve Monges, Stephen~N Thibodeau,
  Roberto Labianca, Stanley~R Hamilton, Amy~J French, Brian Kabat, Nathan~R
  Foster, Valter Torri, et~al.
\newblock Defective mismatch repair as a predictive marker for lack of efficacy
  of fluorouracil-based adjuvant therapy in colon cancer.
\newblock \emph{Journal of Clinical Oncology}, 28\penalty0 (20):\penalty0 3219,
  2010.

\bibitem[Srinidhi et~al.(2021)Srinidhi, Kim, Chen, and
  Martel]{srinidhi2021self}
Chetan~L Srinidhi, Seung~Wook Kim, Fu-Der Chen, and Anne~L Martel.
\newblock Self-supervised driven consistency training for annotation efficient
  histopathology image analysis.
\newblock \emph{arXiv preprint arXiv:2102.03897}, 2021.

\bibitem[Svrcek et~al.(2019)Svrcek, Lascols, Cohen, Collura, Jonch{\`e}re,
  Fl{\'e}jou, Buhard, and Duval]{svrcek2019msi}
Magali Svrcek, Olivier Lascols, Romain Cohen, Ada Collura, Vincent
  Jonch{\`e}re, Jean-Fran{\c{c}}ois Fl{\'e}jou, Olivier Buhard, and Alex Duval.
\newblock Msi/mmr-deficient tumor diagnosis: Which standard for screening and
  for diagnosis? diagnostic modalities for the colon and other sites:
  Differences between tumors.
\newblock \emph{Bulletin du cancer}, 106\penalty0 (2):\penalty0 119--128, 2019.

\bibitem[Yamashita et~al.(2021)Yamashita, Long, Longacre, Peng, Berry, Martin,
  Higgins, Rubin, and Shen]{yamashita2021deep}
Rikiya Yamashita, Jin Long, Teri Longacre, Lan Peng, Gerald Berry, Brock
  Martin, John Higgins, Daniel~L Rubin, and Jeanne Shen.
\newblock Deep learning model for the prediction of microsatellite instability
  in colorectal cancer: a diagnostic study.
\newblock \emph{The Lancet Oncology}, 22\penalty0 (1):\penalty0 132--141, 2021.

\bibitem[Zhang et~al.(2018)Zhang, Osinski, Taxter, Perera, Lau, and
  Khan]{zhang2018adversarial}
Renyu Zhang, Boleslaw~L Osinski, Timothy~J Taxter, Jason Perera, Denise~J Lau,
  and Aly~A Khan.
\newblock Adversarial deep learning for microsatellite instability prediction
  from histopathology slides.
\newblock In \emph{Proceedings of the 1st Conference on Medical Imaging with
  Deep Learning (MIDL 2018), Amsterdam, The Netherlands}, pages 4--6, 2018.

\end{thebibliography}

\end{document}